%% file: multi_objective_pareto_opt.tex
\documentclass[9pt,conference]{IEEEtran}
\pdfoutput=1
\usepackage{graphicx}
\usepackage{epstopdf}
\usepackage{dblfloatfix}
\usepackage{booktabs}
\usepackage{enumitem}
\usepackage{multirow}
\usepackage{url}
\usepackage{breakurl}
\usepackage{subcaption}
\usepackage{color}
\usepackage[font=footnotesize]{caption} 
\usepackage{algorithm}
\usepackage{algorithmic}
\usepackage{amssymb}
\usepackage[cmex10]{amsmath}
\usepackage{amsthm}
\usepackage[numbers,square,comma,sort]{natbib}

\newcommand{\todo}[1]{\textcolor{black}{#1}}
\begin{document}
	
\bstctlcite{IEEEexample:BSTcontrol}




\title{Learning Pareto-Frontier Resource Management Policies for Heterogeneous SoCs: An Information-Theoretic Approach\vspace{-3mm}}


\author{\IEEEauthorblockN{
Aryan Deshwal, Syrine Belakaria, Ganapati Bhat, Janardhan Rao Doppa, Partha Pratim Pande }\vspace{1mm}
	\IEEEauthorblockA{School of Electrical Engineering and Computer Science, 
	Washington State University, Pullman, WA\\
		Email: \{aryan.deshwal, syrine.belakaria, ganapati.bhat, jana.doppa, pande\}@wsu.edu}
}

\IEEEoverridecommandlockouts

\IEEEpubid{\makebox[\columnwidth]{
978-1-6654-3274-0/21/\$31.00~
\copyright2021
IEEE \hfill} \hspace{\columnsep}\makebox[\columnwidth]{ }}
\maketitle

\IEEEdisplaynontitleabstractindextext

\input{files/abstract.tex}

\input{files/introduction.tex}
\input{files/problem_setup.tex}

\input{files/related_work}
\input{files/technical_approach.tex}
\input{files/experimental_evaluation.tex}

\input{files/conclusion_future_research.tex}



\footnotesize{\input{multi_objective_pareto_opt.bbl}}

\end{document}

%% file: files/abstract.tex
\begin{abstract}

Mobile system-on-chips~(SoCs) are growing in their complexity and heterogeneity (e.g., Arm's Big-Little architecture) to meet the needs of emerging applications, including games and artificial intelligence. This makes it very challenging to optimally manage the resources (e.g., controlling the number and frequency of different types of cores) at runtime to meet the desired trade-offs among multiple objectives such as performance and energy. This paper proposes a novel information-theoretic framework referred to as {\em PaRMIS} to create Pareto-optimal resource management policies for given target applications and design objectives. PaRMIS specifies parametric policies to manage resources and learns statistical models from candidate policy evaluation data in the form of target design objective values. The key idea is to select a candidate policy for evaluation in each iteration guided by statistical models that maximize the information gain about the true Pareto front. Experiments on a commercial heterogeneous SoC show that PaRMIS achieves better Pareto fronts and is easily usable to optimize complex objectives (e.g., performance per Watt) when compared to prior methods.

\end{abstract}

%% file: files/introduction.tex
\vspace{-1.5mm}
\section{Introduction}
The usage of mobile platforms and associated applications is growing rapidly \cite{Statista2018_apps}. To meet the needs of emerging applications such as games and artificial intelligence, mobile system-on-chips~(SoCs) are growing in heterogeneity. Emerging heterogeneous mobile SoCs support cores of different types (e.g., Big and Little), and dynamic resource management (DRM) decisions correspond to selecting the number of active cores and corresponding frequency level for each type of core. The DRM problem at runtime as a function of active applications to meet the desired trade-offs among relevant objectives (e.g., performance and energy) poses two key challenges. First, the space of DRM decisions is exponential in the number of cores and their frequencies. For example, Samsung Exynos 5422 SoC has four Big and four Little cores, and we need to select the best decision from 4940 candidates at every decision epoch, which is typically between 50 to 100~ms~\cite{pallipadi2006ondemand}. Second, optimal DRM decisions change depending on the desired trade-off among target objectives. Since required trade-offs can change based on real-world scenarios, we need to create {\em Pareto-frontier} DRM policies to make optimal DRM decisions for different trade-offs. 
The Pareto-frontier allows the selection of a DRM policy that meets the desired trade-off at runtime. 


Prior work on DRM to address the above two challenges is lacking in the following ways. First, solutions in commercial mobile SoCs are based on {\em simple heuristics}. For example, interactive and ondemand governors increase/decrease the operating frequency by one level when the utilization falls below a static threshold. These heuristics only provide a {\em single} trade-off for performance and energy, and they are less accurate than data-driven machine learning based approaches \cite{mandal2019dynamic}. Second, commonly used {\em reinforcement learning (RL)} ~\cite{chen2015distributed,mandal2019dynamic} approaches define a reward function for each objective and consider a linear combination of reward functions with one scalar parameter for each objective.
RL methods learn the DRM policy via trial-and-error based on the feedback from DRM decisions. To create Pareto-frontier DRM policies, we need to run RL methods with different scalar parameters repeatedly.
Unfortunately, accuracy of RL algorithms critically depends on the design of good reward functions, which is hard and even impossible to do for some objectives, e.g., performance per Watt (PPW). Since RL needs to try different actions at each state (i.e., exploration)
to discover the best DRM policy, it may not be feasible for large state space and DRM decision space, as in our case. We may need to run RL with a large number of fine-grained scalar parameter configurations to uncover high-quality Pareto-frontier DRM policies. Third, recent work on supervised approaches to DRM policies is based on the {\em imitation learning (IL)} framework~\cite{mandal2019dynamic}. 
The key idea in IL is to create an Oracle policy for each targeted trade-off and mimic its behavior using off-the-shelf supervised learning algorithms. Recent work has shown the effectiveness of IL for some specific design objectives with minimal trade-off space. Unfortunately, it is computationally hard to create high-quality Oracle policies for complex objectives, such as PPW, for approximating the optimal Pareto front. 



This paper proposes a novel framework referred to as {\em Learning {\bf Pa}reto-frontier {\bf R}esource {\bf M}anagement Policies via {\bf I}nformation-Theoretic {\bf S}earch (PaRMIS)} to automatically create high-quality Pareto-frontier DRM policies for any given set of design objectives, as shown in Figure~\ref{fig:overview}. PaRMIS specifies DRM policy as a function, e.g., multi-layer perceptron~(MLP), with a fixed number of parameters over the system state.
The key idea is to build statistical models over this parameter space by evaluating candidate DRM policies in terms of the given design objectives and using them to select the candidate DRM policy that maximizes the information gain of the optimal Pareto front in each iteration.
We derive an efficient algorithm to compute entropy, a key computational step in the selection procedure. A key feature of our framework is that designers can plug-and-play with any set of target objectives and uncover optimized Pareto-frontier DRM policies in a small number of iterations. Our experimental evaluation on a commercial heterogeneous SoC with 12 applications shows the efficacy and generality of PaRMIS over the state-of-the-art, including interactive and ondemand governors and RL and IL-based methods.

\vspace{1mm}

\noindent {\bf Contributions.} The key contribution is the design, demonstration, and evaluation of the PaRMIS framework to create Pareto-frontier DRM policies for heterogeneous SoCs. {\em To the best of our knowledge, this is the first general framework that directly optimizes for Pareto-frontier DRM policies}. Specific contributions include:

\begin{itemize}
    \item Developing a novel information-theoretic framework referred to as PaRMIS to create resource management policies to trade-off target design objectives such as performance and energy. PaRMIS iteratively selects a candidate policy for evaluation that maximizes the information gain about the optimal Pareto front. 
    \item Development of an efficient algorithm to compute entropy, a key step in the PaRMIS framework.
    \item Comprehensive experiments on a commercial hardware platform using real-world applications to show the advantages of PaRMIS in terms of the quality of Pareto front and ability to optimize complex objectives over state-of-the-art methods. \end{itemize}

%% file: files/problem_setup.tex
\begin{figure}
    \centering
    \vspace{-2mm}
    \includegraphics[width=1\linewidth]{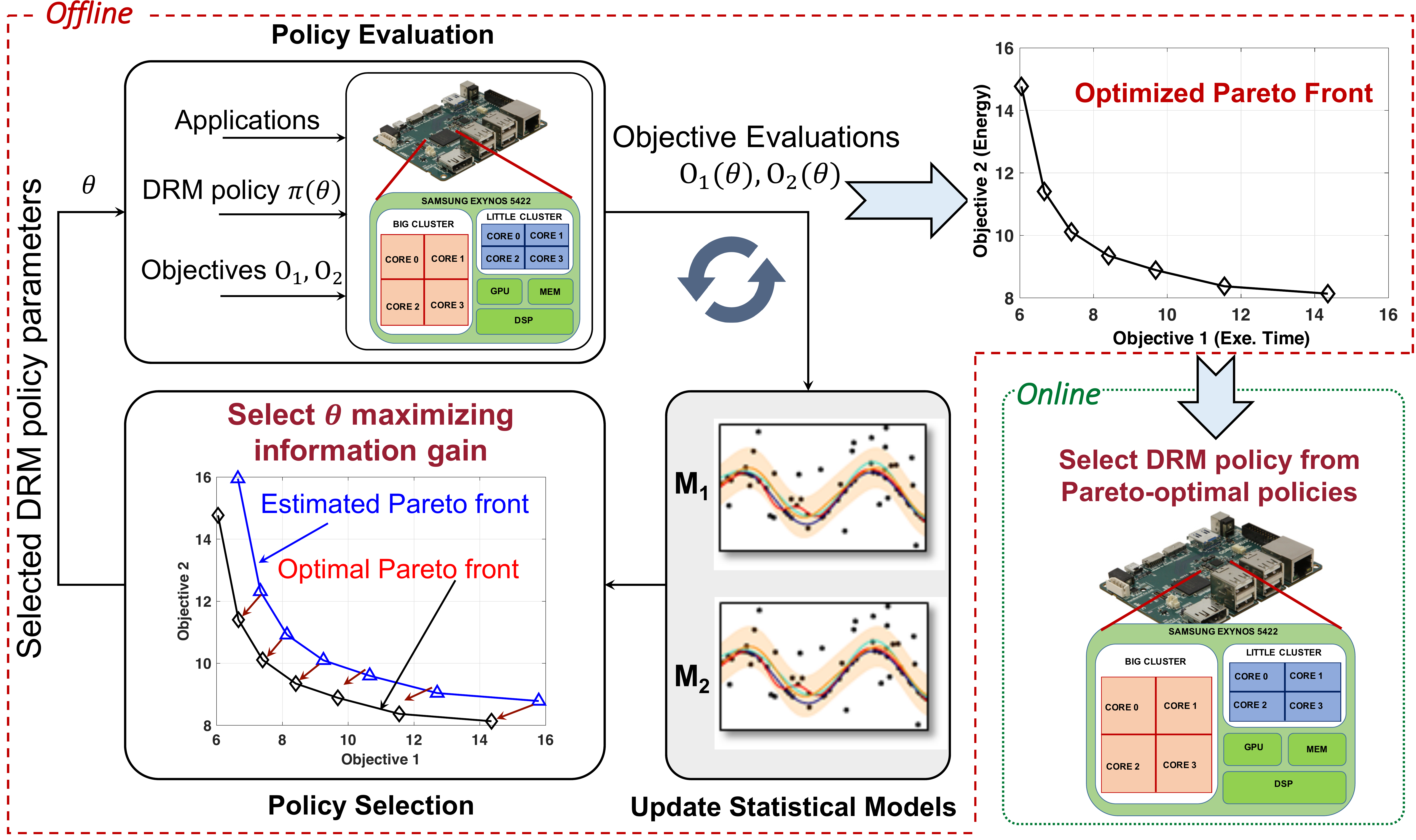}
    \vspace{-4mm}
    \caption{\small{High-level overview of the PaRMIS framework for two objectives. PaRMIS learns two statistical models, one for each objective, using training data in the form of DRM policy parameters $\theta$ (input) and objective evaluations $O_1,O_2$ (output), and uses them to guide the selection of next DRM policy $\Pi_\theta$ that maximizes information gain about the optimal Pareto front. The statistical models are updated with the new training example created in each iteration. At the end of convergence or maximum iterations (offline computation), it produces Pareto-frontier DRM policies: one for each point on the Pareto front. At runtime (online), we select the appropriate DRM policy from this set based on the desired trade-off.}}
    \label{fig:overview}
\end{figure}


\vspace{-0.5mm}
\section{Background and Problem Setup}  \label{sec:problem}
\vspace{-0.5mm}
We consider a heterogeneous mobile platform with $T$ different types of cores and there are $m_i$ number of cores for $i^{th}$ type.
Suppose we can make resource management decisions at {\em runtime} to control the number of active cores and frequencies for each core type. Let each resource management decision be represented by two vectors $(a_1, a_2,\cdots,a_T)$ and $(f_1, f_2,\cdots,f_T)$, where $a_i$ and $f_i$ represent the number of active cores and frequency for $i^{th}$ type, respectively. 
For example, the heterogeneous SoC employed in our experiments has two types of cores: {\em Big $(m_1$ = 4)} and {\em Little $(m_2$ = 4)}. 
Hence, each resource management decision is a four-tuple $(a_{Big}, a_{Little}, f_{Big}, f_{Little})$. Suppose a policy $\Pi$ maps the current system state captured in terms of hardware counters $\Psi$ (see Table~\ref{tb:performance-counters}) to a candidate resource management decision tuple.
\todo{The hardware counters are obtained for a set of repeatable decision epochs for each application. Epochs are clusters of macro-blocks obtained by profiling the basic blocks in an application, as detained in~\cite{gupta2017dypo,mandal2019dynamic}.}
In this work, we consider policies represented as functions with parameters $\theta \in \Re^d$ (e.g., MLP) and consequently denote them as $\Pi_{\theta}$.

Given $k (\geq 2)$ design objectives $\mathcal{O}_1, \mathcal{O}_2,\cdots,\mathcal{O}_k$ (e.g., performance, energy) and a set of applications \textsc{App}, our goal is to create {\em Pareto-frontier} resource management policies $\Pi^*$ to trade-off the given objectives for the application set \textsc{App}. The evaluation of the quality of decisions from a candidate policy $\Pi_{\theta}$ produces a vector of objective values $\vec{O}$ = $(O_1, O_2,\cdots,O_k)$. We say that a policy $\Pi_{\theta}$ {\em Pareto-dominates} another policy $\Pi_{\theta'}$ if $O_i(\Pi_{\theta}) \leq O_i(\Pi_{\theta'}) \hspace{1mm} \forall{i}$ and there exists some $j \in \{1, 2, \cdots,k\}$ such that $O_j(\Pi_{\theta}) < O_j(\Pi_{\theta'})$. The optimal solution of our problem is a set of policies $\Pi^*$ such that no other policy $\Pi_{\theta'}$ Pareto-dominates a policy $\Pi_{\theta} \in \Pi^*$. The solution set $\Pi^*$ is called the optimal {\em Pareto-frontier} resource management policies and the corresponding set of objective values $\mathcal{O}^*$ is called the optimal {\em Pareto front}. 
Once we have a set of Pareto-frontier DRM policies, we select an appropriate policy at runtime based on the desired trade-off among the design objectives.


%% file: files/related_work.tex
\section{Related Work} \label{sec:related_work}
Heterogeneous SoCs are widely used due to their integration of multiple types of cores~(Big/Little), graphics processing units, and other accelerators to support millions of applications~\cite{kumar2005heterogeneous}.
The heterogeneity in the processors necessitates DRM techniques that are able to choose the best configurations as a function of the application requirements~\cite{muthukaruppan2013hierarchical,kadjo2014towards}.
Most DRM techniques, including default governors such as ondemand~\cite{pallipadi2006ondemand}, use core utilization to make their decisions. However, utilization alone does not provide sufficient information about the characteristics of applications running on the system. To address this drawback, recent approaches have used performance counters to make DRM decisions~\cite{aalsaud2016power,park2017ml,reddy2018inter,sartor2020hilite}. 
The performance counters give fine-grained information about the system state, thus allowing DRM policies to make more intelligent decisions. Machine learning approaches, such as decision trees~\cite{park2017ml}, RL~\cite{chen2015distributed}, and IL~\cite{kim2017imitation,mandal2019dynamic,sartor2020hilite} have also been used to create DRM policies for mobile platforms. While these approaches are able to improve upon prior DRM methods, they still optimize for a single objective function, such as energy or execution time or PPW. However, in real-world scenarios, we need DRM policies that can achieve the user's desired trade-off among multiple objectives of interest.
Therefore, there is strong need to develop algorithms to create Pareto-frontier DRM policies so that the system can use the appropriate DRM policy at runtime based on the user's desired trade-off.

\begin{table}[t]
	\centering
	\caption{Features of system state used for DRM policies.}
	\vspace{-4mm}
	\label{tb:performance-counters}
	\begin{tabular}{@{}ll@{}}
		\\ \midrule
		Instructions Retired             & Non-cache External Memory Request      \\
		CPU Cycles                       & Sum of Little Cluster Utilization    \\
		Branch Miss Predictions           & Per Core Big Cluster Utilization \\
		Level 2 Cache Misses             & Total Chip Power Consumption      \\
		Data Memory Accesses               &               \\ \bottomrule
	\end{tabular}
\end{table}

The DyPO approach proposed in~\cite{gupta2017dypo} performs {\em exhaustive search} to find Pareto-frontier points for the objectives of interest and then designs a logistic regression classifier at a {\em coarse-level} over clusters of the Pareto points. Unfortunately, exhaustive search does not scale well with the size of the DRM decision space and number of applications; and the coarse approximation is significantly sub-optimal.
Recent work \cite{kim2017imitation,mandal2019dynamic} used RL and IL to overcome the drawbacks of DyPO by creating DRM policies for a limited number of trade-off scenarios by optimizing a linear combination of the desired objectives. However, RL and IL approaches suffer from two drawbacks. First, they cannot be extended to complex objectives (e.g., PPW) and/or different trade-offs due to the difficulty in designing reward functions and Oracle policies to provide supervision. Second, they don't optimize for Pareto-frontier DRM policies directly and can require significant tuning of scalarization parameters and other hyper-parameters. Third, linear scalarization is known to perform poorly due to its inability to explore non-convex regions of the Pareto front \cite{drawback-linearscal-Das}.
In strong contrast to these approaches, the proposed PaRMIS framework can be used to obtain (near-) optimal Pareto-frontier DRM policies for any given set of design objectives. Experiments on the Odroid-XU3~\cite{ODROID_Platforms} board show that PaRMIS achieves Pareto-fronts that have 13\% and 23\% higher Pareto hypervolume metric compared to state-of-the-art RL and IL methods, respectively.

%% file: files/technical_approach.tex
\section{PaRMIS Framework} \label{sec:overview}


\vspace{1mm}

\noindent {\bf Overview of PaRMIS.} To find optimized Pareto-frontier policies $\Pi^*$, PaRMIS learns statistical models for $k$ design objectives over the parameter space $\theta$ using training data in the form of candidate DRM policy evaluations and iteratively selects the next DRM policy for evaluation. We perform the following algorithmic steps in each iteration: 1) Using the current statistical models, we select the parameters $\theta$ of the candidate DRM policy that maximizes the information gain about the optimal Pareto front $\mathcal{O}^*$. 2) We evaluate the DRM policy $\Pi_{\theta}$ by executing it on the target platform while running the applications to measure the $k$-tuple of objective evaluations $\vec{O}$=$(O_1, O_2,\cdots,O_k)$. 3) We use the new training example in the form of (input) policy parameters $\theta$ and (output) objective evaluations $\vec{O}$=$(O_1, O_2,\cdots,O_k)$ to update the statistical models. At convergence or after maximum number of iterations, we compute the Pareto front from the aggregate set of objective evaluation vectors and output the DRM policies corresponding to Pareto front as the resulting solution. Algorithm \ref{alg:PaRMIS} provides the pseudo-code and Figure~\ref{fig:overview} shows an example illustration of PaRMIS for two design objectives.

\begin{algorithm}[H]
\caption{PaRMIS for Dynamic Resource Management}
\footnotesize
\textbf{Input}: \textsc{Arch} = target heterogeneous SoC, 
\textsc{App} = target applications,
$\mathcal{O}_1,\mathcal{O}_2,\cdots,\mathcal{O}_k$ = $k$ design objectives to trade-off,
$\Pi_{\theta}$ = dynamic resource management policy with parameters $\theta$ \\
\textbf{Output}: Pareto-frontier DRM policies 
\begin{algorithmic}[1]

\STATE Initialize $\mathcal{D}_0 \leftarrow$ initial training data in the form of small number of candidate policy and evaluation objective-vector pairs; and $t \leftarrow$ 0
\REPEAT
\STATE Learn statistical models from training data $\mathcal{D}_t$
\STATE Select candidate DRM  policy parameters that maximize information gain about the true Pareto front: \\ $\theta_{t+1} \leftarrow \arg max_{\theta} \hspace{2 mm} \alpha(\theta) = IG(\{\theta, \vec{O}\}, \mathcal{O}^* \mid \mathcal{D}_t)$ // Eqn. 1 and 9

\STATE Evaluate the selected policy for $k$ objective functions: \\
$\vec{O}$=$({O}_1,{O}_2,\cdots,{O}_k) \leftarrow \textsc{Evaluate}(\textsc{Arch},\textsc{App},\Pi_{\theta_{t+1}})$
\STATE Aggregate the training data: $\mathcal{D}_{t+1} \leftarrow \mathcal{D}_{t} \cup \left\{(\theta_{t+1}, \vec{O})\right\}$
\STATE $t \leftarrow t+1$
\UNTIL{convergence or maximum iterations}
\STATE \textbf{return} the Pareto-frontier DRM policies uncovered during search
\end{algorithmic}
\label{alg:PaRMIS}
\end{algorithm}

\vspace{-3mm}
\subsection{Learning Statistical Models from Training Data}

\noindent {\bf Training data.} We collect training data by iteratively evaluating a sequence of DRM policies. Each training example is of the following form: a) input variables are parameters $\theta$ of the DRM policy $\Pi_\theta$; and b) output variables are objective evaluation vectors  $\vec{O}$=$(O_1, O_2,\cdots,O_k)$ obtained by executing the DRM policy $\Pi_\theta$ when running applications \textsc{App} on the target heterogeneous SoC \textsc{Arch}. Therefore, aggregate training data after $t$ iterations $\mathcal{D}$ consists of $t$ training examples of input-output pairs.

\vspace{1mm}

\noindent {\bf Statistical models.} We want to learn statistical models from training data to capture our uncertainty about the Pareto front and guide us in selecting the candidate DRM policy for evaluation to quickly uncover the Pareto-frontier DRM policies. We employ Gaussian processes (GPs) \cite{williams2006gaussian} as our choice of statistical model due to its superior uncertainty quantification ability via Bayesian interpretation \cite{williams2006gaussian}. A GP over input space $\Theta$ is a random process from $\Theta$ to $\Re$. It is characterized by a mean function $\mu : \Theta  \mapsto \Re$ and a covariance or kernel function $\kappa : \Theta \times \Theta \mapsto \Re$. 
The posterior mean and standard deviation of a GP provide the prediction and uncertainty, respectively. Intuitively, uncertainty will be low for DRM policy parameters $\theta$ that are close to the ones in our training data and will increase as the distance grows. We model the objective functions $\mathcal{O}_1,\mathcal{O}_2,\cdots,\mathcal{O}_k$ using $k$ independent GP models $\mathcal{M}_1,\mathcal{M}_2,\cdots,\mathcal{M}_k$ with zero mean and i.i.d. observation noise and update all statistical models  
from the aggregate training data $\mathcal{D}$ after every iteration.

\subsection{Selecting DRM Policy for Evaluation via Information Gain}

The effectiveness of PaRMIS framework critically depends on the reasoning mechanism to select the candidate DRM policy for evaluation in each iteration. Ideally, we want an algorithmic approach that can use the uncertainty of learned statistical models and allows us to uncover high-quality Pareto-frontier DRM policies in a small number of iterations. Therefore, we propose a novel information-theoretic approach that selects the next candidate DRP policy $\Pi_\theta$ (for ease of notation, we only use parameters $\theta$ in the below discussion) that maximizes the information gain about the {\bf optimal  Pareto front}  $\mathcal{O}^*$. This is equivalent to expected reduction in entropy over the optimal Pareto front $\mathcal{O}^*$. Our utility function that maximizes the information gain between the next candidate input for evaluation $\theta$ and Pareto front $\mathcal{O}^*$ is given as:
\begin{align}
        \alpha(\theta) &= I(\{\theta, \vec{O}\}, \mathcal{O}^* \mid \mathcal{D}) \label{eqn_orig_inf_gain} \\ 
    &= H(\mathcal{O}^* \mid \mathcal{D}) - \mathbb{E}_O [H(\mathcal{O}^* \mid \mathcal{D} \cup \{\theta, \vec{O}\})]  \label{eqn_exp_redn} \\
    &= H(\vec{O} \mid \mathcal{D}, \theta) - \mathbb{E}_{\mathcal{O}^*} [H(\vec{O} \mid D,   \theta, \mathcal{O}^*)] \label{eqn_symmetric_mesmo}
\end{align}
Information gain $I(.)$ is defined as the expected reduction in entropy $H(.)$ of the posterior distribution $P(\mathcal{O}^* \mid \mathcal{D})$ over the optimal Pareto front $\mathcal{O}^*$ as given in Equations \ref{eqn_exp_redn} and \ref{eqn_symmetric_mesmo} resulting from the symmetric property of information gain.

The first term in the r.h.s of Equation \ref{eqn_symmetric_mesmo}, i.e., the entropy of a factorizable $k$-dimensional Gaussian distribution $P(\vec{O}\mid \mathcal{D}, \theta$)) can be computed in closed form as shown below:
\begin{align}
    H(\vec{O} \mid \mathcal{D}, \theta) = \frac{K(1+\ln(2\pi))}{2} +  \sum_{i = 1}^k  \ln (\sigma_i(\theta)) \label{eqn_unconditioned_entropy}
\end{align}
where $\sigma_i^2(\theta)$ is the predictive variance of $i^{th}$ GP model at input $\theta$. Intuitively, it says that the entropy is distributed over the $k$ GP models by the sum of their log standard-deviations. The second term in the r.h.s of equation \ref{eqn_symmetric_mesmo} is an expectation over the Pareto front $\mathcal{O}^*$. We can approximately compute this term via Monte-Carlo sampling as:
\begin{align}
    \mathbb{E}_{\mathcal{O}^*} [H(\vec{O} \mid \mathcal{D},   \theta, \mathcal{O}^*)] \simeq \frac{1}{S} \sum_{s = 1}^S [H(\vec{O} \mid \mathcal{D},   \theta, \mathcal{O}^*_s)] \label{eqn_summation}
\end{align}
where $S$ is the number of samples and $\mathcal{O}^*_s$ denote a sample Pareto front. The main advantages of our utility function are its computational efficiency and accuracy. There are two key algorithmic steps to compute Equation \ref{eqn_summation}, which we describe below:



\noindent {\bf 1) Computing Pareto front samples $\mathcal{O}^*_s$.} To compute a Pareto front sample $\mathcal{O}^*_s$, we first sample functions from the posterior GP models via random Fourier features \cite{random_fourier_features}. Subsequently, we solve a multi-objective optimization over the $k$ sampled functions to capture the interactions between different objectives. We employ the popular NSGA-II algorithm \cite{deb2002nsga} to solve the multi-objective optimization problem with sampled functions noting that any other algorithm can be used to similar effect.

\vspace{1mm}

\noindent {\bf 2) Computing entropy with respect to Pareto front sample $\mathcal{O}^*_s$.}
Let $\mathcal{O}^*_s = \{\vec{z}_1, \cdots, \vec{z}_m \}$ be the sample Pareto front,  where $m$ is the size of the Pareto front and each $\vec{z}_i = \{z_i^1,\cdots,z_i^k\}$ is a $k$-vector evaluated at the $k$ sampled functions. The following inequality holds for each component $O_j$ of the $k$-vector $\vec{O} = \{O_1, \cdots, O_k\}$ in the entropy term $H(\vec{O} \mid \mathcal{D},   \theta, \mathcal{O}^*_s)$:
\begin{align}
 O_j &\leq \max \{z_1^j, \cdots z_m^j \} \quad \forall j \in \{1,\cdots,k\} \label{inequality}
\end{align}

The inequality essentially says that the $j^{th}$ component of $\vec{O}$ (i.e., $O_j$) is upper-bounded by a value obtained by taking the maximum of $j^{th}$ components of all $m$ $k$-vectors in the sample Pareto front $\mathcal{O}^*_s$. This inequality can be proven by a contradiction argument. Suppose there exists some component $O_j$ of $\vec{O}$ such that $ O_j > \max \{z_1^j, \cdots z_m^j \}$. However, by definition, $\vec{O}$ is a non-dominated point because no point dominates it in the $j$th dimension. This results in  $\vec{O} \in \mathcal{O}^*_s$ which is a contradiction. Hence, our hypothesis that $ O_j > \max \{z_1^j, \cdots z_m^j \}$ is incorrect and inequality \ref{inequality} holds.

By combining the inequality \ref{inequality} and the fact that each function is modeled as a GP, we can model each component $O_j$ as a truncated Gaussian distribution since the distribution of $O_j$ needs to satisfy $O_j \leq \max \{z_1^j, \cdots z_m^j \}$.  Furthermore, a common property of entropy measure allows us to decompose the entropy of a set of independent variables into a sum over entropies of individual variables \cite{information_theory}:
\begin{align}
H(\vec{O} \mid \mathcal{D},   \theta, \mathcal{O}^*_s) \simeq \sum_{j=1}^k H(O_j|\mathcal{D}, \theta, \max \{z_1^j, \cdots z_m^j \}) \label{eqn_sep_ineq}
\end{align}

Equation \ref{eqn_sep_ineq} and the fact that the entropy of a truncated Gaussian distribution \cite{entropy_handbook} can be computed in closed form gives the following mathematical expression for the entropy term $H(\vec{O} \mid \mathcal{D},   \theta, \mathcal{O}^*_s)$. 
\begin{align}
    H(\vec{O} \mid \mathcal{D},   \theta, \mathcal{O}^*_s) \simeq & \sum_{j=1}^k [\frac{(1 + \ln(2\pi))}{2}+  \ln(\sigma_j(\theta)) +      \label{eqn_entropy_closed}
 \\ & \ln \Phi(\gamma_s^j(\theta)) - \frac{\gamma_s^j(\theta) \phi(\gamma_s^j(\theta))}{2\Phi(\gamma_s^j(\theta))}] \nonumber
\end{align}
where $\gamma_s^j(\theta) = \frac{O_s^{j*} - \mu_j(\theta)}{\sigma_j(\theta)}$, $y_s^{j*} = \max \{z_1^j, \cdots z_m^j \}$,  and $\phi$ and $\Phi$ are the p.d.f and c.d.f of a standard normal distribution, respectively. By combining equations \ref{eqn_unconditioned_entropy} and \ref{eqn_entropy_closed} with Equation \ref{eqn_symmetric_mesmo}, we get the final form of our utility function as shown below:
\begin{align}
\alpha(\theta) \simeq \frac{1}{S} \sum_{s=1}^S \sum_{j=1}^k \left[ \frac{\gamma_s^j(\theta) \phi(\gamma_s^j(\theta))}{2\Phi(\gamma_s^j(\theta))} - \ln \Phi(\gamma_s^j(\theta)) \right] \label{eqn_final}
 \end{align} 

%% file: files/experimental_evaluation.tex
\section{Experiments and Results} \label{sec:experiments}


\subsection{Experimental Setup} \label{sec:experimental_setup}

\vspace{-0.5mm}

\noindent {\bf Heterogeneous Mobile SoC platform.} We employ the Odroid-XU3 board~\cite{ODROID_Platforms} running Ubuntu 15.04 for our experiments. 
The Exynos 5422 SoC integrates four A15 big cores, four A7 Little cores, a Mali T628 graphics processing unit~(GPU), and other system components. The Odroid board also provides current sensors to measure the power consumption of the big CPU cluster, Little CPU cluster, main memory, and the GPU. We use the on-board current sensors to obtain the energy consumption and PPW metrics to evaluate different DRM policies considered in this paper.


\vspace{0.5mm}

\noindent {\bf Benchmarks.} We employ 12 benchmarks from MiBench~\cite{guthaus2001mibench} and CortexSuite~\cite{thomas2014cortexsuite} suites using the ``large'' input datasets for each suite. 
These benchmarks represent a wide range of real-world scenarios encountered by heterogeneous SoCs.

\vspace{0.5mm}

\noindent {\bf Design objectives.} We consider three objectives, namely, execution time, energy, and PPW to test the generality and effectiveness of different DRM algorithms. 

\vspace{0.5mm}

\noindent {\bf Decision space for DRM policies.} 
For the Odroid-XU3 platform, the decision space is defined by the number of active Big/Little cores and their respective frequencies. There are 4$\times$5 combinations for active cores given that one Little core has to be ON at all times to manage the operating system. Similarly,
the Big and Little core clusters support frequencies from 200 MHz to 2 GHz and 200 MHz to 1.4 GHz in 100 MHz steps, respectively.
Consequently, there are 4$\times$5$\times$13$\times$19 (4940) candidate DRM decisions at each system state. The DRM policy must choose one of these 4940 configurations at each state depending on the desired trade-off among target objectives. 

\vspace{0.5mm}

\noindent \todo{{\bf Decision interval.} An application goes through multiple phases throughout its execution. As a result, using the same configuration for the entire application is not optimal.
To this end, the policies proposed in this paper use the repeatable decision epochs described in~\cite{gupta2017dypo} for making decisions.
Each decision epoch consists of a cluster macro-blocks that capture the varying characteristics of the application.
The policies use the hardware counters (Table~\ref{tb:performance-counters}) observed in each epoch to decide the configuration for the following epoch. 
}

\noindent {\bf Policy representation.} For all learning-based approaches, namely, PaRMIS, RL, and IL, we use one function to make DRM decision for each of the four control knobs at each decision epoch.
In our implementation, we use the following MLP configuration to represent each of the four functions noting that any other function can be used to similar effect: two hidden layers with the ReLU activation and an output layer with the softmax activation. The number of output layer neurons is equal to the number of possible actions for the control knob (e.g., 4 for number of cores). 
We also note that the proposed approach is not dependent on any specific policy representation and other approaches can be used to implement the DRM policies.

\noindent \todo{{\bf Runtime policy selection.} The choice of the DRM policy from the Pareto front depends on the user
preference in terms of desired trade-offs between target objectives, such as power and
performance. For example, if the battery level is low, the user can specify that energy
consumption has the highest priority. 
In this
work, we present our results under the assumption that an interface to provide user
preference about the importance of objectives exists.}

\vspace{-1mm}
\subsection{PaRMIS and Baseline DRM Algorithms}

\noindent {\bf PaRMIS.} There are no critical hyper-parameters to apply PaRMIS. We employed the no. of samples to compute the utility function in Equation \ref{eqn_final} with $S$=1. 
We ran PaRMIS for a maximum of 500 iterations and noticed that it converges in at most 300 iterations.

\vspace{0.5mm}

\noindent \textbf{Reinforcement learning (RL).} 
Prior work using RL has typically focused on optimizing a single objective function by defining an appropriate reward function \cite{chen2015distributed,kim2017imitation,mandal2019dynamic}. Single objective RL algorithms can be extended to multiple objectives by using a linear combination all the objectives via scalarization as: $\mathcal{R}(\mathcal{O}_1,\cdots,\mathcal{O}_k)$ = $\sum_{i=1}^{k} \lambda_i \cdot \mathcal{R}(\mathcal{O}_i)$, where $\mathcal{R}(\mathcal{O}_1,\cdots,\mathcal{O}_k)$ is the combined reward function, and $\mathcal{R}(\mathcal{O}_i)$ and $\lambda_i$ stand for reward function and scalarization parameter for $i$th objective $\mathcal{O}_i$. We employ the reward functions for energy and execution time from recent work \cite{kim2017imitation}. However, it is hard to design a reward function for the PPW objective. We run RL algorithm employed in recent studies \cite{kim2017imitation} with different scalarization parameters $(\lambda_1,\cdots,\lambda_k)$ to create the Pareto-frontier DRM policies.


\vspace{0.5mm}

\noindent \textbf{Imitation learning (IL).} 
IL methods create an Oracle policy to optimize a given objective and then learn a policy to mimic its behavior~\cite{sartor2020hilite,mandal2019dynamic}. We employ the IL approach and Oracle policies for energy and PPW objectives from a recent work \cite{mandal2019dynamic}, which showed good results for optimizing energy with small or no performance penalty (i.e., very specific trade-offs). As noted before, Oracle policies may not be optimal for some objectives such as PPW and for different trade-offs. Similar to RL, we run IL by creating Oracle policies to optimize a linear combination of target objectives and obtain Pareto-frontier DRM policies by varying the scalarization parameters.


\vspace{0.5mm}

\noindent\textbf{Default governors.} We also compare with the default governors in the system, i.e., ondemand, interactive, performance, and powersave. These governors provide a single point on the Pareto-front since they are optimized for a single objective, such as power or performance. Nonetheless, it is crucial to compare with these baselines as they are implemented on millions of commercial platforms.


\vspace{-0.4ex}

\subsection{Quality of Application-Specific Pareto-front}
\vspace{-0.4ex}
In this section, we compare different DRM algorithms for each application separately for two objectives: execution time and energy.
To this end, we compute Pareto-frontier DRM policies using PaRMIS, RL, and IL approaches by running them on a single application and measure the quality of the resulting Pareto-front. These results provide the best-case scenario for each application and help us in analyzing how global DRM policies learned over all applications compare to application-specific DRM~policies.

\vspace{0.5mm}

\noindent {\bf PHV metric.} We employ the Pareto hypervolume (PHV) metric, which is commonly used to measure the quality of a given Pareto front \cite{zitzler1999evolutionary}. PHV is defined as the volume between a reference point and the given Pareto front. We report the normalized PHV metric w.r.t the PHV of PaRMIS approach (higher the better).

\vspace{0.5mm}

\begin{figure}[t]
    \centering
    \includegraphics[width=0.95\linewidth]{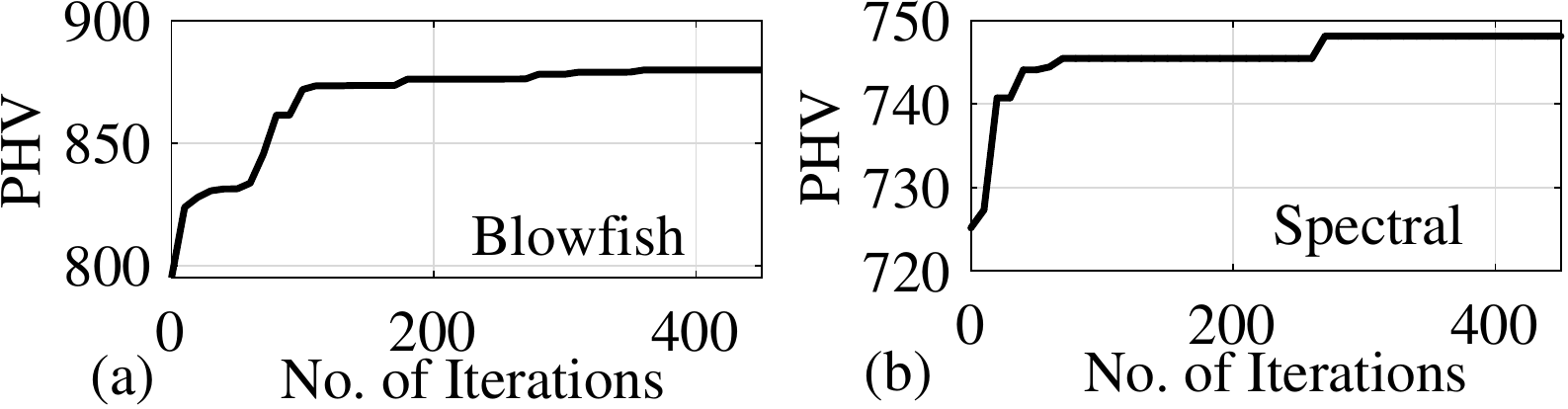}
    \caption{Convergence of PaRMIS for (a) Blowfish and (b) Spectral.}
    \vspace{-0.5mm}
    \label{fig:convergence}
\end{figure}
\noindent {\bf Convergence of PaRMIS.} Recall that PaRMIS is an iterative approach and we want to see the number of iterations required to converge to the uncovered Pareto-front.
Figure \ref{fig:convergence} shows PHV of the Pareto-front vs. no. of iterations for Blowfish and Spectral benchmarks noting that other applications show similar or better convergence behavior. We can see that PHV improvement is significant in the initial iterations and converges in at most 300 iterations. 

\begin{figure}[t]
    \centering
    \includegraphics[width=1\linewidth]{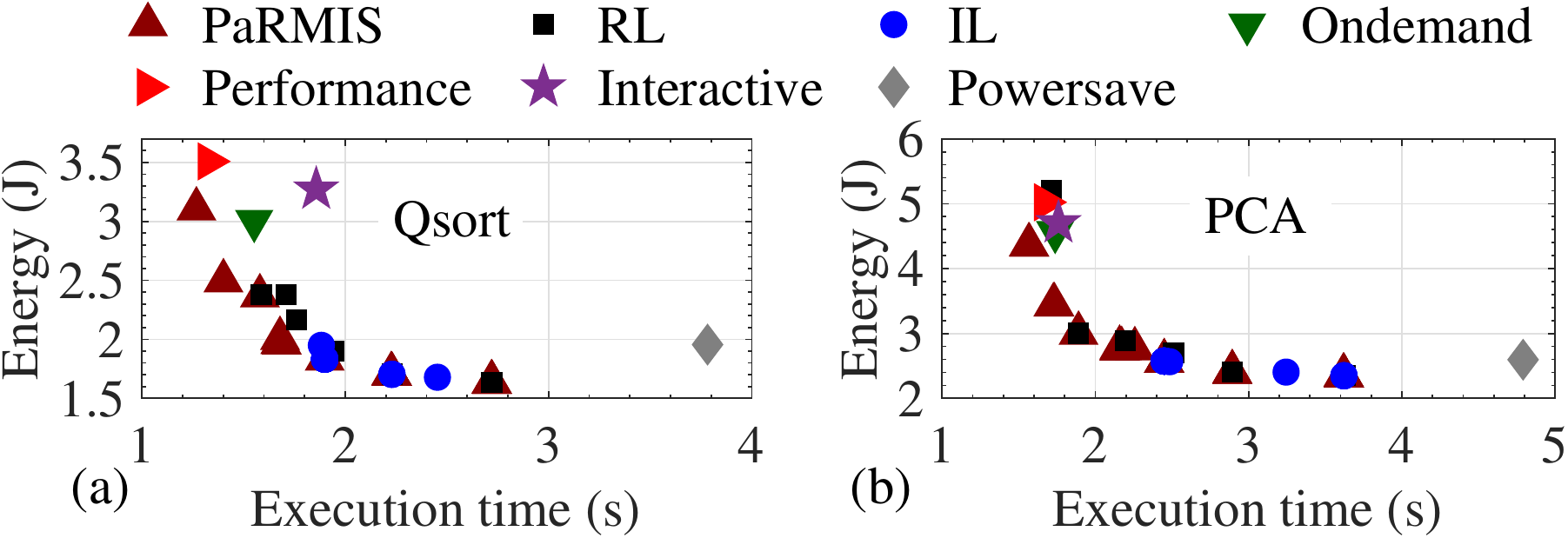}
    \caption{Application-specific Pareto-front for (a) Qsort and (b) PCA.}
    \label{fig:energy_app_pareto}
    \vspace{2mm}
    \includegraphics[width=0.96\linewidth]{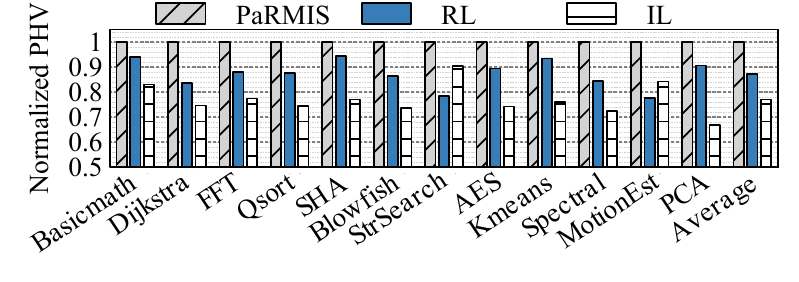}
    \vspace{-0.5mm}
    \caption{Comparison of normalized PHV metric of baseline methods w.r.t the PaRMIS approach for application-specific optimization.}
    \label{fig:PHV_comparison_app}
    \vspace{-0.5mm}
\end{figure}

\noindent {\bf Energy consumption vs. Execution time Pareto front.} Figure~\ref{fig:energy_app_pareto} shows the overall Pareto-front for two representative benchmarks (Qsort and PCA) noting that we got similar results for all applications. Each marker in the figure corresponds to one policy from the Pareto-frontier DRM policy set obtained by PaRMIS (dark red $\triangle$), RL (black $\square$), and IL (blue $\circ$), respectively. We make the following observations. 1) The Pareto-front obtained by PaRMIS dominates those from both RL and IL. More specifically, PaRMIS creates DRM policies that improve both objectives when compared to RL and IL. Furthermore, PaRMIS creates policies that have a wider range of trade-offs between energy and execution time. For example, the lowest execution time obtained by PaRMIS for the Qsort application is 1.2~s, while the lowest values for RL and IL are 1.6~s and 1.9~s, respectively. 2)  Figure~\ref{fig:energy_app_pareto} also shows the trade-off obtained by DRM policies of the four default governors. We can clearly see that the Pareto-front obtained by PaRMIS dominates all of them significantly. The difference is especially visible for the performance governor that is optimized for minimizing the execution time. Even in this case, PaRMIS is able to provide a DRM policy that has both lower execution time and energy than the performance governor. In summary, these results show that PaRMIS creates DRM policies that provide significant improvements over both the default governors and state-of-the-art machine-learning based DRM approaches.

\vspace{0.5mm}

\vspace{0.5mm}

\noindent {\bf PHV comparison.} The data in Figure~\ref{fig:energy_app_pareto} offers an intuitive visualization of the Pareto-fronts obtained by each DRM approach. However, it does not allow a quantitative comparison of the Pareto-front quality.
The PHV metric allows us to compare the quality of different Pareto-fronts. For computing PHV, the reference point is chosen such that it has a higher execution time and energy than all points in the Pareto front. To allow comparison between different Pareto-fronts, the same reference point is used for all DRM approaches. Figure~\ref{fig:PHV_comparison_app} shows the comparison of normalized PHV metric for all the 12 applications.
The normalized PHV of both RL and IL is significantly lower than 1, which shows that they have a significantly lower PHV. 
For example, PaRMIS has 10\% and 25\% higher PHV than RL and IL for the PCA application. On an average, PaRMIS achieves 13\% and 23\% higher PHV compared to RL and IL, respectively. This shows that the quality of the Pareto front obtained by PaRMIS is consistently better than both RL and IL.
Prior work \cite{mandal2019dynamic,kim2017imitation} has shown that IL is better than RL for specific trade-offs between energy and execution time.
However, IL performs worse than RL over the entire Pareto-front because the Oracle policy for different trade-offs is not optimal. 
RL and IL also suffer from drawbacks of linear scalarization  due to its inability to explore non-convex regions of the Pareto front \cite{drawback-linearscal-Das}. These results show the key advantage of PaRMIS not requiring any effort from designers to get the optimized Pareto-front.

\begin{figure}[b]
    \centering
    \includegraphics[width=0.96\linewidth]{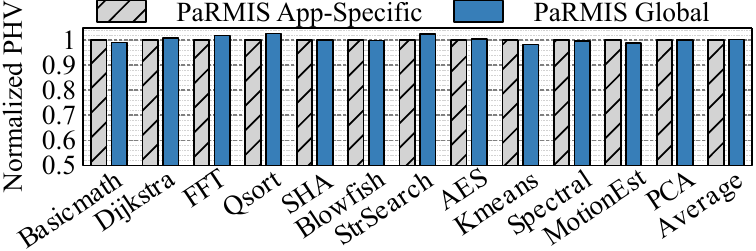}
    \caption{Comparison of normalized PHV of PaRMIS for application-specific vs. global Pareto-frontier DRM policies.}
    \label{fig:phv_global}
\end{figure}

\subsection{Global vs. Application-Specific Pareto-frontier DRM policies}
Application-specific policies do not scale as the number of applications available to the user grow in size. Moreover, not all applications are known at design-time. Therefore, DRM algorithms must learn {\em global Pareto-frontier DRM policies} that are applicable to all the applications. To this end, we apply PaRMIS to design global Pareto-frontier policies using 
training data from all 12 applications.

%

Figure~\ref{fig:phv_global} shows the normalized PHV for all the applications. The PHV is normalized with respect to the PHV of application-specific Pareto-front. As expected, the normalized PHV of the global Pareto-frontier policies is within 2\% of the application-specific  policies. For FFT, Qsort, and StringSearch, the PHV of the global Pareto-frontier policies is higher than the application-specific Pareto-frontier policies. On an average, the PHV of global and application-specific Pareto-frontier policies are equal. In summary, the global Pareto-frontier policies achieve comparable or better quality than application-specific  policies while generalizing to all applications. 


\subsection{Evaluation with Complex Objectives}
\begin{figure}[t]
    \centering
    \includegraphics[width=1\linewidth]{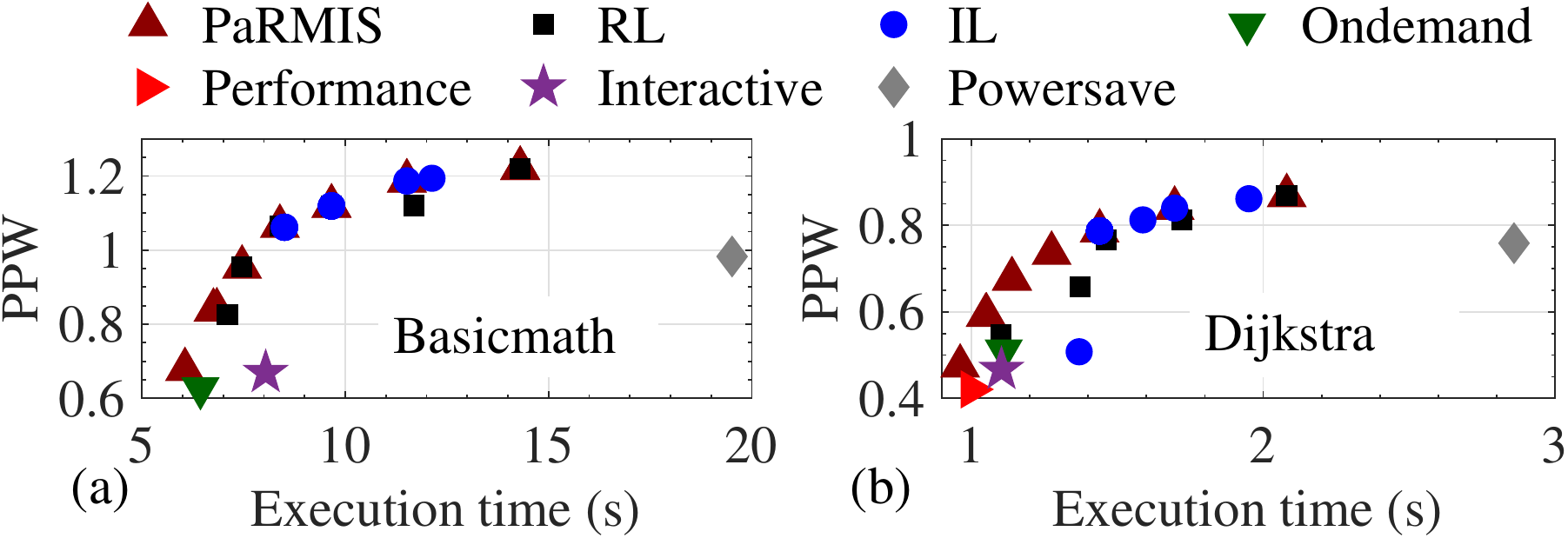}
    \caption{Comparison of application-specific Pareto-front to optimize PPW and execution time for (a) Basicmath and (b) Dijkstra.}
    \label{fig:pareto_ppw}
    \vspace{2mm}

    \includegraphics[width=0.96\linewidth]{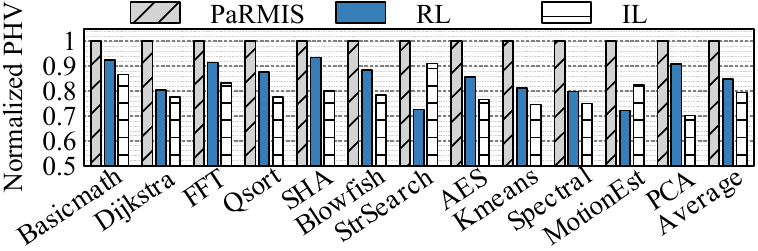}
    \vspace{-0.5mm}
    \caption{Normalized PHV metric of baseline methods w.r.t the PaRMIS approach for application-specific optimization over PPW and execution time. 
    }
    \label{fig:phv_ppw}
    \vspace{-1mm}
\end{figure}

One of the main advantages of the PaRMIS approach is that it can be easily applied with any set of complex objectives desired by the designers. Recall that this is not possible with RL and IL as it is hard to design a good reward function and optimal Oracle policy respectively for complex objectives such as PPW. To demonstrate this advantage, we use PaRMIS to optimize PPW and execution time for each application. However, we cannot use RL and IL to optimize PPW and execution time as PPW is a complex, non-linear objective. There is no reward function and optimal Oracle policy for PPW objective~\cite{mandal2020energy}. 
Due to these limitations, we reuse the Pareto-frontier DRM policies for energy and execution time from RL and IL, and compute the Pareto-front and PHV for PPW and execution time objectives.
Figure~\ref{fig:pareto_ppw} shows a comparison of the Pareto fronts obtained by PaRMIS, RL, and IL for Basicmath and Dijkstra applications. The Pareto front achieved by PaRMIS dominates those from RL and IL both in terms of the range of policies and quality of individual Pareto points. PaRMIS is also able to dominate the default governors available on the platform. A similar behavior is seen for the normalized PHV metric, as shown in Figure~\ref{fig:phv_ppw}. PaRMIS has a higher PHV for all the applications with an average improvement of 16\% and 21\% over RL and IL, respectively. These results show that PaRMIS can be easily extended to any new and complex objectives.

\begin{table}[b]
\footnotesize
\centering
\vspace{1mm}
\caption{Summary of implementation overhead.}
\label{tab:overhead}
\vspace{-1mm}
\begin{tabular}{@{}lrrrr@{}}
\toprule
Metric    & Per Policy & Total  & \% Overhead \\ \midrule
Exe. time & 200 $\mu$s & 800 $\mu$s                                                       & 0.8 (every 100 ms)         \\
Memory    & 1 KB       & 27 KB                                                        & 0.001       \\ \bottomrule
\end{tabular}
\end{table}

\subsection{Implementation Overhead}
The DRM policies in each approach are implemented as user-space governors in software to characterize the overhead.
Furthermore, all learning-based approaches including PaRMIS, RL, and IL use the same MLP function with different set of parameters to represent each DRM policy in the user-space governor. Hence, the storage cost and decision-making time for each policy is same for all three methods. 
In particular, contrary to existing implementation that employs look up table for RL~\cite{kim2017imitation}, we use the same function approximator to implement both RL and IL. Hence, there is no computational and storage difference between IL, RL, and PaRMIS.
Table~\ref{tab:overhead} provides a summary of all overheads.
On an average, per decision execution of a DRM policy to choose the runtime configuration takes about 800~$\mu$s (200~$\mu$s for each knob), which amounts to about 0.8\% overhead when DRM decisions are made every 100~ms.
The memory required to store a single DRM policy from all three methods (PaRMIS, RL, and IL) is 1 KB. When we employ global Pareto-frontier policies, PaRMIS creates 27 policies that form the Pareto front, resulting in 27 KB storage overhead (0.001\% with 2 GB RAM available on the SoC platform). At runtime, we choose one DRM policies from this set of 27 policies as per the desired trade-off. In summary, the overhead in terms of storage and DRM decision-making time is negligible.

%% file: files/conclusion_future_research.tex
\vspace{-0.5mm}
\section{Conclusions and Future Work} \label{sec:conclusion}
\vspace{-0.5mm}
Dynamic resource management (DRM) of mobile SoCs is a challenging problem due to rise of heterogeneity, large state space and decision space, and complexity of application workloads. This paper studied a novel information-theoretic learning framework referred to as PaRMIS to create Pareto-frontier DRM policies. PaRMIS can produce high-quality DRM policies and easy to configure/apply to trade-off any set of complex design objectives.
Experiments on a commercial heterogeneous SoC platform show that PaRMIS achieves Pareto-fronts that have 13\% and 23\% higher Pareto hypervolume (PHV) compared to state-of-the-art RL and IL methods, respectively. 
Immediate future work includes studying PaRMIS for large-scale manycore systems.

\vspace{0.5ex}

\noindent {\bf Acknowledgements.} This work was supported in part by the NSF grants CNS-1955353, OAC-1910213 and IIS-1845922, in part by the ARO grants W911NF-17-1-0485 and W911NF-19-1- 0162, and in part by semiconductor research corporation's AI Hardware program.

\vspace{-0.5mm}

%% file: multi_objective_pareto_opt.bbl

%% file: multi_objective_pareto_opt.bbl
\begin{thebibliography}{10}
\providecommand{\url}[1]{#1}
\csname url@samestyle\endcsname
\providecommand{\newblock}{\relax}
\providecommand{\bibinfo}[2]{#2}
\providecommand{\BIBentrySTDinterwordspacing}{\spaceskip=0pt\relax}
\providecommand{\BIBentryALTinterwordstretchfactor}{4}
\providecommand{\BIBentryALTinterwordspacing}{\spaceskip=\fontdimen2\font plus
\BIBentryALTinterwordstretchfactor\fontdimen3\font minus
  \fontdimen4\font\relax}
\providecommand{\BIBforeignlanguage}[2]{{%
\expandafter\ifx\csname l@#1\endcsname\relax
\typeout{** WARNING: IEEEtranS.bst: No hyphenation pattern has been}%
\typeout{** loaded for the language `#1'. Using the pattern for}%
\typeout{** the default language instead.}%
\else
\language=\csname l@#1\endcsname
\fi
#2}}
\providecommand{\BIBdecl}{\relax}
\BIBdecl

\bibitem{aalsaud2016power}
A.~Aalsaud \emph{et~al.}, ``{Power--Aware Performance Adaptation of Concurrent
  Applications In Heterogeneous Many-Core Systems},'' in \emph{ISLPED}, 2016.

\bibitem{chen2015distributed}
Z.~Chen \emph{et~al.}, ``{Distributed Reinforcement Learning For Power Limited
  Many-Core System Performance Optimization},'' in \emph{DATE}, 2015.

\bibitem{information_theory}
T.~M. Cover and J.~A. Thomas, \emph{{Elements of Information Theory}}, 2012.

\bibitem{drawback-linearscal-Das}
I.~Das \emph{et~al.}, ``{A Closer Look at Drawbacks of Minimizing Weighted Sums
  of Objectives for Pareto Set Generation in Multicriteria Optimization
  Problems},'' \emph{Structural optimization}, vol.~14, no.~1, pp. 63--69,
  1997.

\bibitem{deb2002nsga}
K.~Deb \emph{et~al.}, ``{A Fast and Elitist Multiobjective Genetic Algorithm:
  NSGA-II},'' \emph{IEEE TEC}, vol.~6, no.~2, pp. 182--197, 2002.

\bibitem{gupta2017dypo}
U.~Gupta \emph{et~al.}, ``{DyPO: Dynamic Pareto-Optimal Configuration Selection
  for Heterogeneous MpSoCs},'' \emph{ACM TECS}, 2017.

\bibitem{guthaus2001mibench}
M.~R. Guthaus \emph{et~al.}, ``{MiBench: A Free, Commercially Representative
  Embedded Benchmark Suite},'' in \emph{Proc. WWC-4}, 2001, pp. 3--14.

\bibitem{ODROID_Platforms}
{Hardkernel}. (2014) Odroid-xu3.
  \url{https://www.hardkernel.com/shop/odroid-xu3/} Accessed 11/20/2020.

\bibitem{kadjo2014towards}
D.~Kadjo \emph{et~al.}, ``{Towards Platform Level Power Management In Mobile
  Systems},'' in \emph{Int. Syst.-on-Chip Conf. (SOCC)}, 2014, pp. 146--151.

\bibitem{kim2017imitation}
R.~Kim \emph{et~al.}, ``{Imitation Learning For Dynamic VFI Control In
  Large-Scale Manycore Systems},'' \emph{IEEE TVLSI}, vol.~25, no.~9, 2017.

\bibitem{kumar2005heterogeneous}
R.~Kumar \emph{et~al.}, ``{Heterogeneous Chip Multiprocessors},''
  \emph{Computer}, vol.~38, no.~11, pp. 32--38, 2005.

\bibitem{mandal2019dynamic}
S.~K. Mandal \emph{et~al.}, ``{Dynamic Resource Management of Heterogeneous
  Mobile Platforms via Imitation Learning},'' \emph{IEEE TVLSI}, 2019.

\bibitem{mandal2020energy}
S.~K. Mandal \emph{et~al.}, ``{An Energy-Aware Online Learning Framework for
  Resource Management in Heterogeneous Platforms},'' \emph{ACM TODAES},
  vol.~25, no.~3, pp. 1--26, 2020.

\bibitem{entropy_handbook}
J.~V. Michalowicz, J.~M. Nichols, and F.~Bucholtz, \emph{Handbook of
  differential entropy}.\hskip 1em plus 0.5em minus 0.4em\relax Chapman and
  Hall/CRC, 2013.

\bibitem{muthukaruppan2013hierarchical}
T.~S. Muthukaruppan \emph{et~al.}, ``{Hierarchical Power Management For
  Asymmetric Multi-Core In Dark Silicon Era},'' in \emph{DAC}, 2013.

\bibitem{pallipadi2006ondemand}
V.~Pallipadi and A.~Starikovskiy, ``{The Ondemand Governor},'' in \emph{Proc.
  Linux Symp.}, vol.~2, 2006, pp. 215--230.

\bibitem{park2017ml}
J.-G. Park \emph{et~al.}, ``{ML-Gov: A Machine Learning Enhanced Integrated
  CPU-GPU DVFS Governor For Mobile Gaming},'' in \emph{Proc. of ESTIMedia},
  2017, pp. 12--21.

\bibitem{random_fourier_features}
A.~Rahimi and B.~Recht, ``{Random Features for Large-scale Kernel Machines},''
  in \emph{NeurIPS}, 2008, pp. 1177--1184.

\bibitem{reddy2018inter}
B.~K. Reddy \emph{et~al.}, ``{Inter-cluster Thread-to-core Mapping and DVFS on
  Heterogeneous Multi-cores},'' \emph{IEEE TVLSI}, vol.~4, no.~3, 2018.

\bibitem{sartor2020hilite}
A.~Sartor \emph{et~al.}, ``{HiLITE: Hierarchical and Lightweight Imitation
  Learning for Power Management of Embedded SoCs},'' \emph{IEEE CAL}, vol.~19,
  no.~1, pp. 63--67, 2020.

\bibitem{Statista2018_apps}
Statista, ``{Mobile App Usage - Statistics \& Facts},''
  \url{https://www.statista.com/topics/1002/mobile-app-usage/} Accessed 24 Nov.
  2018.

\bibitem{thomas2014cortexsuite}
S.~Thomas \emph{et~al.}, ``{CortexSuite: A Synthetic Brain Benchmark Suite.}''
  in \emph{IISWC}, 2014, pp. 76--79.

\bibitem{williams2006gaussian}
C.~K. Williams and C.~E. Rasmussen, \emph{Gaussian processes for machine
  learning}.\hskip 1em plus 0.5em minus 0.4em\relax MIT Press, 2006, vol.~2,
  no.~3.

\bibitem{zitzler1999evolutionary}
E.~Zitzler, \emph{{Evolutionary Algorithms for Multiobjective Optimization:
  Methods and Applications}}, 1999, vol.~63.

\end{thebibliography}
